\documentclass[twocolumn,pre,showpacs]{revtex4}
\usepackage[dvips]{graphicx}

\begin{document}

\title{The internal modes of sine-Gordon solitons in the presence of
spatiotemporal perturbations}
\author{J. A. Gonz\'{a}lez}
\affiliation{Laboratorio de F\'{\i}sica Computacional, Centro de F\'{\i}sica, 
Instituto Venezolano de Investigations Cient\'{\i}ficas,
Apartado 21827, Caracas 1020-A, Venezuela}
\author{A. Bellor\'{\i}n}
\affiliation{Laboratorio de F\'{\i}sica Te\'orica de S\'olidos, Escuela de F\'{\i}sica,
Facultad de Ciencias, Universidad Central de Venezuela,
Apartado 47586, Caracas 1041-A, Venezuela}
\author{L. E. Guerrero}
\affiliation{Grupo de la Materia Condensada, Departamento de F\'{\i}sica, 
Universidad Sim\'{o}n Bol\'{\i}var,
Apartado 89000, Caracas 1080-A, Venezuela}
\date{\today}

\begin{abstract}
\noindent We investigate the dynamics of the sine-Gordon solitons perturbed
by spatiotemporal external forces. We prove the existence of internal
(shape) modes of sine-Gordon solitons when they are in the presence of
inhomogeneous space-dependent external forces, provided some conditions (for
these forces) hold. Additional periodic time-dependent forces can sustain
oscillations of the soliton width. We show that, in some cases, the internal
mode even can become unstable, causing the soliton to decay in an
antisoliton and two solitons. In general, in the presence of spatiotemporal
forces the soliton behaves as a deformable (non-rigid) object. A soliton
moving in an array of inhomogeneities can also present sustained
oscillations of its width. There are very important phenomena (like the
soliton-antisoliton collisions) where the existence of internal modes plays
a crucial role.
\end{abstract}

\preprint{HEP/123-qed}
\maketitle

The sine-Gordon solitons are very important in physics. They possess crucial
applications in both particle physics and condensed matter theory. For
instance, in solid state physics, they describe domain walls in
ferromagnets, dislocations in crystals, charge density waves, fluxons in
long Josephson junctions and Josephson transmission lines, etc. \cite%
{Lonngren,Bishop,Kivshar,Scott,McLaughlin,Braun}

In general, nonintegrable soliton equations (e.g. the $\varphi ^{4}$
equation and the double sine-Gordon) may possess internal degrees of
freedom, which are crucial in many phenomena \cite%
{Peyrard,Campbell,Campbell2,Kivshar2}. A recent discussion of internal modes
of solitary waves can be found in Ref. \cite{Kivshar3}.
However, it is well-known that the unperturbed (``pure'') sine-Gordon
equation does not have internal modes.

A very remarkable question is the following: \emph{can external forces
create internal modes in the sine-Gordon equation?}

Recently there has been a hot debate in the scientific literature about the
existence of internal modes of sine-Gordon solitons.
Some authors \cite{Rice,Boesch,Tchafo,Zhang,Zhang2,Zhang3,Majernikova} have
claimed that they have found an internal quasimode described as a long-lived
oscillation of the width of the sine-Gordon soliton.

On the other hand, a very recent and interesting paper is contradicting all
these reports \cite{Quintero}.
By considering the response of the soliton to ac forces and initial
distortions, Quintero \textit{et al }show that neither intrinsic internal
modes nor ``quasimodes'' exist in contrast to previous reports. We should
stress that they use only time-dependent perturbations in their work.

In the present Rapid Communication we study the sine-Gordon equation
perturbed by spatiotemporal external forces: 
\begin{equation}
\phi _{tt}+\gamma \phi _{t}-\phi _{xx}+\sin \phi =F(x,t).  \label{1}
\end{equation}

We will show that with some spatially inhomogeneous forces, the internal
modes can exist for the sine-Gordon equation.

We have shown in previous papers \cite%
{Gonzalez,Gonzalez2,Guerrero,Gonzalez3,Gonzalez4,Gonzalez5,Gonzalez6} that
in equations as the following: 
\begin{equation}
\phi _{tt}+\gamma \phi _{t}-\phi _{xx}+\sin \phi =F_{1}(x),  \label{2}
\end{equation}%
if the force $F_{1}(x)$ possesses a zero $x^{*}$ ($F_{1}(x^{*})=0$), this
can be an equilibrium position for the soliton. If there is only one zero,
this is a stable equilibrium position for the soliton if $\left( \frac{%
\partial F_{1}(x)}{\partial x}\right) _{x^{*}}>0$. For the antisoliton, it
is stable if $\left( \frac{\partial F_{1}(x)}{\partial x}\right) _{x^{*}}<0$.

Let us suppose that $F_{1}(x)$ is defined as: 
\begin{equation}
F_{1}(x)=2(B^{2}-1)\sinh (Bx)/\cosh ^{2}(Bx).  \label{3}
\end{equation}
This is a function with a zero in the point $x^{*}=0$.

We have chosen this function because of the following properties: (i) the
exact solution for the soliton resting on the equilibrium position can be
obtained, and (ii) the stability problem of this soliton can be solved
exactly. The results obtained with this function can be generalized
qualitatively to other systems topologically equivalent to this one.
Besides, real
physical systems are related to this example \cite{Kivshar}. For instance,
in a Josephson junction a perturbation that can be described by a function
of type $F(x)=\frac{dR(x)}{dx}$, where $R(x)$ is a bell-shaped function, is
an Abrikosov vortex lying in the junction's plane perpendicular to its local
dimension \cite{Aslamasov}.

A similar function can describe a local deformation of a charge density wave
system \cite{Brasovsky}.

Usually, function $R(x)$ is taken as the Dirac's $\delta $-function.
However, if we wish to model a finite-width inhomogeneity, function (\ref{3}%
) is a better choice.

The exact stationary solution of Eq. (\ref{2}), with $F_{1}(x)$ as defined
in (\ref{3}), is 
$\phi _{k}=4\arctan \left[ \exp \left( Bx\right) \right]$ .

The stability analysis, which considers small amplitude oscillations around $%
\phi _{k}$ $\left[ \phi (k,x)=\phi _{k}(x)+f(x)e^{\lambda t}\right] $, leads
to the eigenvalue problem \cite%
{Gonzalez,Gonzalez2,Guerrero,Gonzalez3,Gonzalez4,Gonzalez5,Gonzalez6}: 
$\widehat{L}f=\Gamma f$, 
where $\widehat{L}=-\partial _{x}^{2}+\left[ 1-2\cosh ^{-2}(Bx)\right] $ and 
$\Gamma =-\lambda ^{2}-\gamma \lambda $.

This problem can be solved exactly \cite{Flugge}. The eigenvalues of the
discrete spectrum \cite%
{Gonzalez,Gonzalez2,Guerrero,Gonzalez3,Gonzalez4,Gonzalez5,Gonzalez6} are
given by the formula 
\begin{equation}
\Gamma _{n}=B^{2}(\Lambda +2\Lambda n-n^{2})-1,  \label{6}
\end{equation}%
where $\Lambda (\Lambda +1)=2/B^{2}$.

The integer part of $\Lambda $, i.e. $\left[ \Lambda \right] $, yields the
number of eigenvalues in the discrete spectrum, which correspond to the
soliton modes (this includes the translational mode $\Gamma _{0}$, and the
internal or shape modes $\Gamma _{n}$ with $n>0$ \cite%
{Gonzalez,Gonzalez2,Guerrero,Gonzalez3,Gonzalez4,Gonzalez5,Gonzalez6}.

All this theoretical investigation produces the following results (note that
parameter $B$ will be our control parameter):
For $B^{2}>1$, the translational mode is stable and there are no internal
modes.
If $\frac{1}{3}<B^{2}<1$, then the translational mode is unstable. However,
still there are no internal modes.
When $\frac{1}{6}<B^{2}<\frac{1}{3}$, apart of the translational mode, there
is one internal mode! This internal mode is stable.
In the case that $B^{2}<\frac{1}{6}$
there can appear many other internal modes! The exact number
is $\left[ \Lambda \right] -1$, where $\Lambda (\Lambda +1)=2/B^{2}$.
For $B^{2}<\frac{2}{\Lambda _{*}(\Lambda _{*} +1)}$, where 
$\Lambda _{*}= \frac{5+\sqrt{17}}{2}$, the first internal mode
becomes unstable!

\emph{What happens when we shift the soliton center-of-mass away from the
equilibrium position?}

We have the following initial problem: 
\begin{equation}
\phi (x,0)=4\arctan \left\{ \exp \left[ B\left( x-x_{0}\right) \right]
\right\} ,  \label{7}
\end{equation}%
\begin{equation}
\phi _{t}(x,0)=0.  \label{8}
\end{equation}

In the stable case ($B^{2}>1$) the center-of-mass of the soliton will make
damped oscillations (for $x_{0}\neq 0$) around the equilibrium point $x=0$.

In the case that the translational mode is unstable ($\frac{1}{3}< B^{2}<1 $%
), the soliton will move away indefinitely from the equilibrium position.

Consider the next initial problem: 
\begin{equation}
\phi (x,0)=4\arctan \left[ \exp \left( Bx\right) \right] +C\sinh (Bx)\cosh
^{-\Lambda }(Bx),  \label{9}
\end{equation}%
\begin{equation}
\phi _{t}(x,0)=0.  \label{10}
\end{equation}

In this initial problem the initial soliton is deformed.

\begin{figure}[tbh]
\centerline{\includegraphics[width=3in]{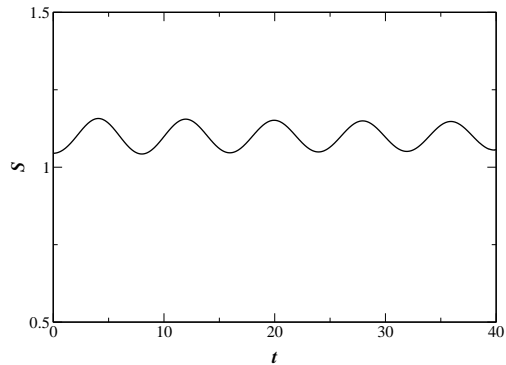}}
\caption{Soliton's width oscillations when the internal mode can be
excited and it is stable, $\frac{1}{6}<B^{2}<\frac{1}{3}$.}
\label{fig3}
\end{figure}

For $\frac{1}{6}<B^{2}<\frac{1}{3}$ we will observe oscillations of the
soliton width (see Fig.~\ref{fig3}). This is due to the fact that an
internal mode has been excited. Eventually, due to unavoidable errors in the
initial conditions or to energy exchange between the internal mode and the
translational mode, the soliton will move away from the equilibrium position
(remember that the equilibrium position is unstable for the soliton
center-of-mass). 

It is important to note that the instability of the translational mode does
not mean instability of the soliton structure.

We have to say that the frequency of the oscillations observed in the
numerical simulations coincides with the one obtained theoretically using
equation (\ref{6}). The frequency of the width oscillations can be obtained
using the equations $\omega _{1}=\sqrt{\Gamma _{1}}$, where $\Gamma
_{1}=B^{2}(3\Lambda -1)-1$. All our experiments confirm the prediction
about the frequency of the shape oscillations.

\begin{figure}[tbh]
\centerline{\includegraphics[width=3in]{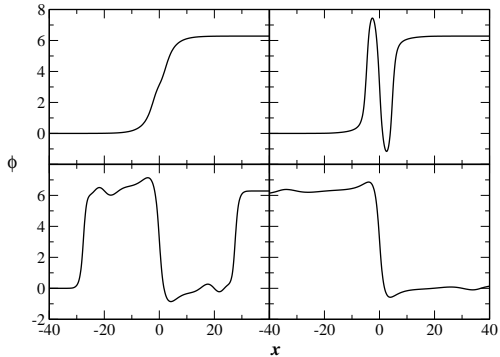}}
\caption{Soliton's destruction when the internal mode is unstable,
$B^{2}<\frac{2}{\Lambda _{*}(\Lambda _{*} +1)}$, 
where
$\Lambda _{*}= \frac{5+\sqrt{17}}{2}$.}
\label{fig4}
\end{figure}

The most spectacular phenomenon occurs for 
$B^{2}<\frac{2}{\Lambda (\Lambda +1)}$, 
$\Lambda = \frac{5+\sqrt{17}}{2}$.
In this
case, the first internal mode is unstable. If we study the evolution of the
soliton from the initial conditions (\ref{9})-(\ref{10}) we will observe the
destruction of the soliton (see Fig.~\ref{fig4}). Two solitons move away (in
different directions) to ``infinity'' (or to the boundaries of the system
and an antisoliton is formed in the place of the original soliton remaining
there stabilized. In fact, the condition $B^{2}<1$ implies stability for the
center-of-mass of an antisoliton. 

Note that in these situations, the sine-Gordon solitons do not behave as
rigid objects, which is what is expected from them in general \cite%
{Quintero2}.
The initial distortions of the width of the soliton will eventually be
damped due to dissipation.

Once the internal modes are possible, as in Eqs. (\ref{2})-(\ref{3}) with $%
\frac{1}{6}<B^{2}<\frac{1}{3}$, we need time-dependent external forces to
sustain the oscillations of the soliton width.

On the other hand, if we wish the soliton to remain in some spatially
localized zone, we need stable equilibrium positions for the center-of-mass
of the soliton.

Let us consider the following spatiotemporal perturbation: 
\begin{equation}
\phi _{tt}+\gamma \phi _{t}-\phi _{xx}+\sin \phi =F_{2}(x)+F_{3}(x,t),
\label{11}
\end{equation}%
where

\[
F_{2}(x)= \left\{ 
\begin{array}{lcl}
F_{1}(x) & , & \mathrm{if}\;-x_{1} \le x \le x_{1}, \\ 
{\displaystyle \frac{A}{\cosh \left[ B\left( x+x_{1}\right) \right] }-D} & ,
& \mathrm{if}\;x<-x_{1}, \\ 
{\displaystyle D-\frac{A}{\cosh \left[ B\left( x-x_{1}\right) \right] }} & ,
& \mathrm{if}\;x>x_{1},%
\end{array}%
\right. 
\]
and $F_{3}(x,t)=f_0\cos(\omega t)[1/\mathrm{cosh}^2(E(x+x_1)) + 1/\mathrm{cosh}^2(E(x-x_1))$.
The space-dependent force $F_{2}(x)$ creates a double-well potential for the
soliton.
At the same time, in the interval $-x_{1}<x<x_{1}$, we have the same force $%
F_{1}(x)$, which was sufficient for the existence of the internal mode.

Actually, other forces can be used to excite the internal mode. In fact, if
we have a function $F(x)$ that can mimic approximately the behavior of
function $F_{1}(x)$ (specially in the interval $-x_{1}<x<x_{1}$, where $%
-x_{1}$ and $x_{1}$ are the extrema of function $F_{1}(x)$) when $B$
satisfies\ the condition $\frac{1}{6}<B^{2}<\frac{1}{3}$, then this function
is good for exciting the internal mode. And note that the behavior of
function $F_{1}(x)$ in the interval $-x_{1}<x<x_{1}$ is a very common
behavior for a function in an interval where there is a zero and two
extrema. 

The time-dependent force $F_{3}(x,t)$ will cause the soliton width to
oscillate. The center-of-mass of the soliton will also oscillate, jumping
between the potential wells created by force $F_{2}(x)$.

Although the soliton is not always in the interval $-x_{1}<x<x_{1}$, it will
return to this interval regularly. While the soliton is in this interval,
all the conditions hold for the internal mode to be excited.

\begin{figure}[!hbt]
\centerline{%
\includegraphics[width=3in]{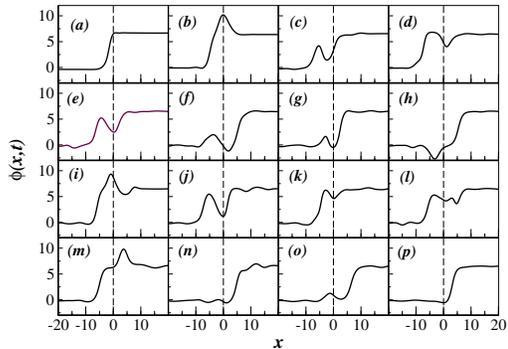}}
\caption{Soliton profiles for different time instants governed
by Eq. (\ref{11})
($B=0.5,\;D=0.2,\;\gamma=0.1,\;f_0=0.7,\;\omega=0.55,\;E=0.7$).
}
\label{fig5a}
\end{figure}

Figure~\ref{fig5a} shows the extraordinary deformations
suffered by the soliton in some cases. 

However, we have corroborated
that other spatiotemporal forces also can sustain the soliton width
oscillations. This includes the ``ubiquitous'' force $F_{3}(t)=f_{0}\cos(%
\omega\,t)$.

The propagation of solitons in disordered media has been studied intensively
in recent years \cite{Gonzalez6,Gredeskul}.

Consider the equation 
\begin{equation}
\phi _{tt}+\gamma \phi _{t}-\phi _{xx}+\sin \phi =F(x).  \label{12}
\end{equation}%
where $F(x)$ is defined in such a way that it possesses many zeroes, maxima
and minima. This system describes an array of
inhomogeneities.

For our study, we have defined $F(x)$ in the following way: 
\begin{equation}
F(x)= {\displaystyle \sum _{n=-q}^{q} 4\left(1-B^2\right) \frac{\mathrm{e}%
^{B\left(x+x_n\right)}-\mathrm{e}^{3B\left(x+x_n\right)}}{\left(\mathrm{e}%
^{2B\left(x+x_n\right)}+1\right)^2}}  \label{13}
\end{equation}
where $x_n = (n+2)\log\left(\sqrt{2}+1\right)/B$ ($n=-q,-q+1\cdots ,q-1,q$),
and $q+2$ is the number of extrema points of $F(x)$. 

In our array, there is a ``superposition'' of the ``disorder'' with a dc
component which will cause the soliton to move to the right all the time.
When the soliton is moving over  intervals where $\frac{dF(x)}{dx}<0$, the
internal mode can be excited. In fact, the points $x_{i}$ where $F(x_{i})=0$
and $\frac{dF(x_{i})}{dx}<0$, are ``barriers'' which the soliton can
overcome due to its kinetic energy. These ``collisions'' with the barriers
will excite the internal modes if in these intervals the function $F(x)$
mimics the behavior of $F_{1}(x)$ when $\frac{1}{6}<B^{2}<\frac{1}{3}$.
The simulations show that the width of the soliton will perform
sustained oscillations during its motion in a disordered medium. 

We have shown that in sine-Gordon equations perturbed by inhomogeneous
(space-dependent) forces $F(x)$, the solitons can possess internal modes.

Some of our results are in agreement with previous works \cite%
{Quintero,Quintero2}. In fact, in Eq. (\ref{2}), with $F_{1}(x)$ as in (\ref%
{3}), if we put $B^{2}=1$, then there are no external forces, and from Eq. (%
\ref{6}) we obtain that there are no internal modes in that case neither.

Moreover, even when there is an inhomogeneous external force, not for every $%
F(x)$ we have internal modes.

For instance, if $F(x)$ has a zero which corresponds to a stable equilibrium
position for the soliton, even then the internal modes are impossible. This
explains why it has been so difficult to find sine-Gordon internal modes.
For the existence of internal modes for sine-Gordon solitons we need zeroes
of function $F(x)$ that corresponds to unstable positions for the
center-of-mass of the soliton.
When the soliton center-of-mass is very close to an unstable equilibrium
position, there is a pair of forces acting \ in opposite directions on the
``body'' of the soliton. This pair of forces should be sufficiently large to
stretch the soliton ``body'', such that the soliton internal mode can be
excited.

Function $F(x)$ can possess several zeroes corresponding to unstable and
stable equilibrium positions. For instance, we have studied a force $F(x)$
that creates a double-well potential for the soliton.

Periodic time-dependent forces (besides the spatially inhomogeneous forces)
can sustain the oscillations of the soliton width.

A soliton moving in an array of inhomogeneities can also undergo sustained
oscillations of its width.

All this is possible because the internal mode of the soliton can exist when
it is moving in media where there are inhomogeneous space-dependent forces
with unstable equilibrium positions.

Nonetheless, we have discovered another more remarkable phenomenon: The
sine-Gordon internal mode not only can exist for some external forces, but
(in some situations) it can become unstable.
If we have an unstable equilibrium position for the soliton center-of-mass
and the pair of forces acting on the soliton is too large, then the soliton
can be destroyed. 

When the soliton is destroyed, it can be transformed into an antisoliton and
two new solitons. The topological charge is conserved. We had found this
phenomenon before for the $\phi ^{4}$-equation \cite{Gonzalez2,Guerrero}.
However, here for the first time we have shown not only that the sine-Gordon
soliton internal mode can exist, but that it can become unstable and destroy
the soliton.
This is a spectacular manifestation of the fact that the sine-Gordon soliton
can behave as a deformable (non-rigid) object.

A. Bellor\'{\i}n would like to thank CDCH-UCV for support under
project PI-03-11-4647-2000.

\end{document}